\documentclass[review]{elsarticle}

\usepackage[margin=1in]{geometry}
\usepackage{float}
\usepackage{lineno,hyperref}
\usepackage{color}
\usepackage{amssymb}
\usepackage{graphics}
\usepackage{graphicx}
\usepackage{pbox}
\usepackage{mhchem}
\usepackage{xcolor} 
\usepackage[ampersand]{easylist}
\usepackage{lineno}
\usepackage{booktabs}
\usepackage{multirow}
\usepackage{subcaption}
\usepackage{amsmath}
\usepackage{paralist}
\usepackage{soul}
\usepackage{siunitx}
\usepackage{amsmath}
\usepackage{url}
\usepackage{makecell}
\usepackage{hyperref}
\usepackage{subcaption}
\usepackage{booktabs}

\newcommand{\argmin}{\mathop{\mathrm{argmin}}\limits}
\newcommand{\solve}{\mathop{\text{Solve}}\limits}
\newcommand{\cm}{\text{cm}} 
\newcommand{\g}{\text{g}}

\newcommand{\MeV}{\text{MeV}} 

\newcommand{\dep}{\text{dep}}
\newcommand{\detector}{\text{det}}
\newcommand{\en}{\text{en}}

\newcommand{\PE}{\text{PE}}
\newcommand{\CS}{\text{CS}}
\newcommand{\PP}{\text{PP}}

\hypersetup{  
    colorlinks=true,
    linkcolor=blue,
    filecolor=magenta,      
    urlcolor=cyan,
    pdftitle={Sharelatex Example},
    bookmarks=true,
    pdfpagemode=FullScreen,
    citecolor=blue
    }

\modulolinenumbers[1]

\makeatletter
\def\ps@pprintTitle{%
 \let\@oddhead\@empty
 \let\@evenhead\@empty
 \def\@oddfoot{\centerline{\thepage}}%
 \let\@evenfoot\@oddfoot}
\makeatother

\begin{document}

\begin{frontmatter}

\title{A Semiempirical Transparency Model for Dual Energy Cargo Radiography Applications}

\author[MITaddress]{Peter Lalor\corref{corauthor}}
\cortext[corauthor]{Corresponding author 
\\\hspace*{13pt} Email address: plalor@mit.edu
\\\hspace*{13pt} Telephone: (925) 453-1876 
\\\hspace*{13pt} 138 Cherry St, Cambridge, MA 02139}

\author[MITaddress]{Areg Danagoulian}

\address[MITaddress]{Department of Nuclear Science and Engineering, Massachusetts Institute of Technology, Cambridge, MA 02139, USA}

\begin{abstract}
Cargo containers passing through ports are scanned by non-intrusive inspection systems to search for concealed illicit materials. By using two photon beams with different energy spectra, dual energy inspection systems are sensitive to both the area density and the atomic number of cargo contents. Most literature on the subject assumes a simple exponential attenuation model for photon intensity in which only free streaming photons are detected. However, this approximation neglects second order effects such as scattering, leading to a biased model and thus incorrect material predictions. This work studies the accuracy of the free streaming model by comparing it to simulation outputs, finding that the model shows poor atomic number reconstruction accuracy at high-$Z$ and suffers significantly if the source energy spectra and detector response function are not known exactly. To address these challenges, this work introduces a semiempirical transparency model which modifies the free streaming model by rescaling different components of the mass attenuation coefficient, allowing the model to capture secondary effects ignored by the free streaming model. The semiempirical model displays improvement agreement with simulated results at high-$Z$ and shows excellent extrapolation to materials and thicknesses which were not included during the calibration step. Furthermore, this work demonstrates that the semiempirical model yields accurate atomic number predictions even when the source spectra and detector response are not known exactly. Using the semiempirical model, manufacturers can perform a simple calibration to enable more precise $Z$ reconstruction capabilities, which has the potential to significantly improve the performance of existing radiographic systems.
\end{abstract}
\end{frontmatter}


\section{Introduction}
\label{Introduction}

In 2021, the Customs and Border Protection (CBP) processed more than 32.7 million imported cargo containers through U.S. ports of entry~\cite{CBP2021}. A principle concern is that a terrorist might attempt to smuggle a nuclear or radiological dispersal device through a port and subsequently detonate a weapon on U.S. soil. Such an attack could have devastating economic consequences in excess of \$1 trillion due to infrastructure damage, trade disruption, and port shutdown costs~\cite{Meade2006, Rosoff2007}. The international atomic energy agency (IAEA) tracks incidents of nuclear and other radioactive material out of regulatory control, identifying 320 incidents of trafficking or malicious use since 1993~\cite{IAEA2022}. For these reasons, U.S. congress passed the SAFE Port Act in 2006, mandating 100 percent screening of U.S. bound cargo and 100 percent scanning of high-risk containers~\cite{PLAW109-347}.

When a cargo container passes through a U.S. port, it is first passively screened by a radiation portal monitor (RPM). This systems detect neutron and gamma radiation which is being passively emitted from radiological sources which may be hidden inside a cargo container~\cite{Kouzes2005, Kouzes2008}. However, a smuggler could evade passive detection through sufficient shielding of the radiological device~\cite{Gaukler2010, Miller2015}. As a result, radiography systems are also utilized to complement passive detection~\cite{NII}. These X-ray and gamma ray imaging systems produce a sensitive density image of the scanned cargo, enabling the identification of nuclear threats which have been shielded to avoid passive detection. Furthermore, through the use of dual energy radiography, these systems can produce a rough elemental analysis of cargo contents, since the attenuation of photons is dependent on the atomic number of the material. This enhances the capabilities of radiography systems to detect nuclear threats and high-$Z$ shielding materials~\cite{Miller2015}.

\section{Background}
\label{Background}

\subsection{Dual Energy Radiography Overview}

When a radiography system scans a material of area density $\lambda$ and atomic number $Z$, it measures a transparency $T(\lambda, Z)$, defined as the detected charge in the presence of the material normalized by the open beam measurement:

\begin{equation}
T(\lambda, Z) =  \frac{Q(\lambda, Z)}{Q_{\text{air}}}
\label{transparency}
\end{equation}

The exact functional form of Eq. \ref{transparency} is arbitrarily complex, since it depends on the precise geometric details of any particular scanning system. Importantly, Eq. \ref{transparency} depends on the atomic number of the imaged object through the mass attenuation coefficient $\mu(E, Z)$, which describes the attenuation of photons through matter by the Beer-Lambert law:

\begin{equation}
\frac{I}{I_0} = e^{-\mu(E, Z) \lambda}
\label{BeerLambert}
\end{equation}

where $E$ is the photon energy and $I/I_0$ is the ratio of transmitted to initial photon intensity. The dual energy principle observes that the mass attenuation coefficient depends on both energy and atomic number, and thus  multiple transparency measurements of the same material using different high and a low energy beam spectra may allow for a determination of the atomic number $Z$ of the imaged object. We use the subscript $H$ to refer to the high energy beam, and the subscript $L$ to refer to the low energy beam. As such, for a pair of measured transparencies $\{T_H, T_L\}$, the corresponding area density and atomic number estimates $\hat \lambda$ and $\hat Z$ are determined by solving the following $2 \times 2$ system:

\begin{equation}
\hat \lambda, \hat Z = \solve\limits_{\lambda, Z}
\begin{cases} 
      T_H(\lambda, Z) = T_H \\
      T_L(\lambda, Z) = T_L
\end{cases}
\label{2x2eq}
\end{equation}

Older works attempted to solve Eq. \ref{2x2eq} directly~\cite{Alvarez1976, Novikov1999}. However, this approach is generally too slow given the millions of pixels in a radiograph image and often only yields approximate solutions. Instead, Eq. \ref{2x2eq} is generally solved through the use of reverse lookup tables~\cite{Zhang2005}. In other words, for a $(\lambda, Z)$ range of interest, $T_H(\lambda, Z)$ and $T_L(\lambda, Z)$ are evaluated and tabulated. Then, for a pair of transparency measurements $\{T_H, T_L\}$ of an unknown material, $\{\lambda, Z\}$ are reconstructed by reference to the lookup table. This approach is fast, accurate, and conveniently simple.

The key remaining ingredient needed to solve Eq. \ref{2x2eq} is an accurate choice of transparency models $T_H(\lambda, Z)$ and $T_L(\lambda, Z)$. To do this, existing dual energy literature can be broadly classified into two categories: analytic methods and empirical methods. Analytic models are derived from the exponential dependence of photon intensity (Eq. \ref{BeerLambert}), which neglects scattered photons and thus introduces a significant source of error. Empirical models typically require a burdensome calibration step and show limited atomic number selectivity for materials not included in the calibration. This work introduces a hybrid method, which addresses both of these challenges by defining a semiempirical transparency model.

\subsection{Analytic Methods}
\label{Analytic Methods}

Most authors define a transparency model by constructing a simple, closed form expression for $T_{\{H, L\}}(\lambda, Z)$~\cite{Novikov1999, Zhang2005, Ogorodnikov2002, Li2016}. Eq. \ref{freestreaming} shows the typical choice, which we call the free streaming transparency model:

\begin{align}
\begin{split}
&T_H(\lambda, Z)_{\text{free streaming}} = \frac{\int_0^{\infty} D(E) \phi_H (E) e^{-\mu (E, Z) \lambda} dE}{\int_0^{\infty} D(E) \phi_H (E) dE} \\
&T_L(\lambda, Z)_{\text{free streaming}} = \frac{\int_0^{\infty} D(E) \phi_L (E) e^{-\mu (E, Z) \lambda} dE}{\int_0^{\infty} D(E) \phi_L (E) dE}
\end{split}
\label{freestreaming}
\end{align}

where $\phi_{\{H, L\}} (E)$ are the $\{$high, low$\}$ energy differential photon beam spectra, and $D(E)$ is the detector response function, calculated as

\begin{equation}
D(E) = C \int_0^{E} R(E, E_\dep) E_\dep dE_\dep
\label{q_Ephoton}
\end{equation}

where $R(E, E_\dep)$ is the differential detector response matrix, representing the probability that a photon with incident energy $E$ deposits energy $E_\dep$, and $C$ is a proportionality constant. Through direct evaluations of Eq. \ref{freestreaming}, a large, high-resolution lookup table can be constructed, enabling atomic number discrimination~\cite{Zhang2005}. Other works perform a variable transformation instead of using $T_{\{H, L\}}(\lambda, Z)$ directly, and construct lookup tables on these new features~\cite{Novikov1999, Ogorodnikov2002, Li2016}. These methods are all expected to give similar results.

A fundamental assumption made by the free streaming model is that only noninteracting photons will be detected and thus ignores the effects of scattered radiation. Past works have claimed that these approximations were validated experimentally and through the use of Monte Carlo simulations~\cite{Zhang2005, Ogorodnikov2002}. Detector systems are typically heavily collimated to reduce the significance of scattered radiation, although it can sometimes be as much as one percent of the primary detector beam~\cite{Chen2007}. Section \ref{Analysis} uses Monte Carlo simulations to analyze the accuracy of the free streaming model, finding that the model yields inaccurate atomic number reconstruction at high-$Z$. Furthermore, the free streaming model shows significant bias if the beam spectra and detector response are not known exactly. These results highlights the primary disadvantage of analytic methods, as a biased transparency model will yield inaccurate atomic number predictions.

\subsection{Empirical Methods}
\label{Empirical Methods}

To circumvent the biases introduced by the free streaming model, some authors instead construct the transparency model empirically through the use of calibration measurements~\cite{Budner2006}. To do this, transparencies of a known material are measured for a range of different thicknesses, and the results are tabulated. This is repeated for different materials to construct a lookup table, and the $\{\lambda, Z\}$ of an unknown material are again determined by reference to the lookup table. This is the most common approach used by commercial dual energy cargo inspection systems.

The primary advantage of empirical approaches is the high degree of accuracy. Since the transparency model is constructed directly from the data, there is no inherent model bias. However, the data acquisition step can often be long and laborious. Empirical methods require a significant quantity of high-precision measurements to construct the lookup tables. Subsequently, if a target material $\{\lambda, Z\}$ does not exist in the lookup table, an interpolation scheme is necessary to predict the identity of the material, introducing a source of error. Typically, only a few calibration materials are used, and the task is simplified to discrimination between organics, organics-inorganics, inorganics, and heavy substances~\cite{Ogorodnikov2002}. However, such a coarse label discretization reduces accuracy and can yield high variance results if a target material is near the midpoint of two bins.

Lee et al. employs a different empirical approach by expressing a material-selective projection image as a linear combination of polynomial basis functions~\cite{Lee2012, Lee2018}. Then, for a choice of basis materials, a set of material-specific weighting factors are calculated through a least squares minimization against a calibration phantom. After the system has been calibrated, material classification is performed by evaluating the material-selective projections, which indicate whether the corresponding basis material is present in the image. The authors show that this approach can yield accurate material discrimination capabilities without knowledge of the photon energy spectra or the detector response function. However, this method is incapable of identifying materials with atomic numbers that differ significantly from the basis materials, resulting in limited atomic number selectivity.

\subsection{A Semiempirical Transparency Model}
\label{Semiempirical method}

To address the challenges described in sections \ref{Analytic Methods} and \ref{Empirical Methods}, this work introduces a semiempirical transparency model. First, we define a semiempirical mass attenuation coefficient, $\tilde \mu(E, Z; a, b, c)$, as follows:

\begin{equation}
\tilde \mu(E, Z; a, b, c) = a\mu_\PE(E, Z) + b\mu_\CS(E, Z) + c\mu_\PP(E, Z)
\label{semiempirical_mass_atten}
\end{equation}

where $\mu_\PE(E, Z)$, $\mu_\CS(E, Z)$, and $\mu_\PP(E, Z)$ are the mass attenuation coefficients from the photoelectric effect (PE), Compton scattering (CS), and pair production (PP), respectively. Values for the mass attenuation coefficients are calculated from NIST cross section tables~\cite{NIST}. In Eq. \ref{semiempirical_mass_atten}, $a$, $b$, and $c$ are calibration parameters which scale the relative importance of photoelectric absorption, Compton scattering, and pair production on the net photon attenuation. In the limit $a=b=c=1$, the nominal mass attenuation coefficient is recovered. The motivation behind this approach is that by rescaling different components of the mass attenuation coefficient, it may be possible to capture some of the secondary effects ignored by the free streaming model. The semiempirical transparency model is then adapted from the free streaming model (Eq. \ref{freestreaming}) as follows:

\begin{align}
\begin{split}
&\tilde T_H(\lambda, Z; a_H, b_H, c_H) = \frac{\int_0^{\infty} D(E) \phi_H (E) e^{-\tilde \mu (E, Z; a_H, b_H, c_H) \lambda} dE}{\int_0^{\infty}D(E) \phi_H (E) dE} \\
&\tilde T_L(\lambda, Z; a_L, b_L, c_L) = \frac{\int_0^{\infty} D(E) \phi_L (E) e^{-\tilde \mu (E, Z; a_L, b_L, c_L) \lambda} dE}{\int_0^{\infty}D(E) \phi_L (E) dE}
\end{split}
\label{semiempirical}
\end{align}

As to be described in section \ref{Analysis}, the semiempirical transparency model shows improved agreement with simulated data compared to the free streaming model, particularly when the source energy spectra and detector response function are not known exactly. Additionally, the semiempirical transparency model requires far less calibration data than fully empirical methods and shows excellent extrapolation to elements which were not included in the calibration step.

\section{Analysis}
\label{Analysis}

\subsection{Accuracy of the Free Streaming Transparency Model}
\label{accuracy_freestreaming}

In order to quantify the accuracy of the free streaming model (Eq. \ref{freestreaming}), transparency measurements were simulated in Geant4~\cite{Geant4,grasshopper}. In each simulation, dual energy bremsstrahlung beams with endpoint energies of $10 \MeV$ and $6 \MeV$ were directed through targets composed of different materials and thicknesses and measured by a stack of collimated cadmium tungstate (CdWO$_4$) scintillators. This outputs a simulated set of transparency data points to compare to the model predictions. The simulation geometry is described in more detail in section \ref{Simulation Details}.

Fig. \ref{T_H_lambda_freestreaming} compares the simulated transparency measurements to the free streaming model predictions, revealing an obvious model bias. To analyze the atomic number discrimination capabilities of the model, Fig. \ref{alpha_curve_freestreaming} shows the simulated measurements and the model predictions on an $\alpha$-curve. An $\alpha$-curve is a plot of $\alpha_H - \alpha_L$ versus $\alpha_H$ for different elements and area densities, where a log transform $\alpha \rightarrow -\log T$ has been performed. Every element on an $\alpha$-curve forms a characteristic $\alpha$-line, describing the relative effect on $T_H$ and $T_L$ as area density is varied. An $\alpha$-curve is a useful visualization tool because slight transparency differences between the high and low energy measurements become more apparent, and the separation between different $\alpha$-lines offers insights into atomic number discrimination capabilities. Fig. \ref{alpha_curve_freestreaming} reveals that the free streaming model accurately reconstructs the $\alpha$-lines of low- to med-$Z$ materials, but becomes less accurate at high-$Z$.

\subsection{Accuracy of the Semiempirical Transparency Model}
\label{accuracy_semiempirical}

First, $a$, $b$, and $c$ are calibrated by minimizing the squared logarithmic error between the semiempirical model (Eq. \ref{semiempirical}) and simulated calibration measurements:

\begin{align}
\begin{split}
&a_H, b_H, c_H = \argmin_{a, b, c} \sum_i \left(\log \tilde T_H(\lambda_i, Z_i; a, b, c) - \log T_{H,i}\right)^2 \\
&a_L, b_L, c_L = \argmin_{a, b, c} \sum_i \left(\log \tilde T_L(\lambda_i, Z_i; a, b, c) - \log T_{L,i}\right)^2
\end{split}
\label{calc_abc}
\end{align}

In Eq. \ref{calc_abc}, the summation index $i$ labels the simulation of calibration material $\{\lambda_i, Z_i\}$ with corresponding output transparencies $\{T_{H,i}, T_{L,i}\}$. In this work, carbon $(Z=6)$, iron $(Z=26)$, and lead $(Z=82)$ were chosen as calibration materials due to their availability in real applications and because they span a wide range of $Z$. High resolution simulations were performed at an area density of $\lambda = 150 \g/\cm^2$ to generate the calibration dataset. At least three calibration materials are required, and this work finds that thicker materials yield a more accurate calibration. The calibration parameters are determined separately for the high and low energy beams. The best fit values of $a$, $b$, and $c$ are shown in the first two rows of table \ref{abc_fit} for each incident beam spectrum.

\begin{figure}
    \centering
    \begin{subfigure}[t]{0.48\textwidth}
        \centering
        \includegraphics[width=\textwidth]{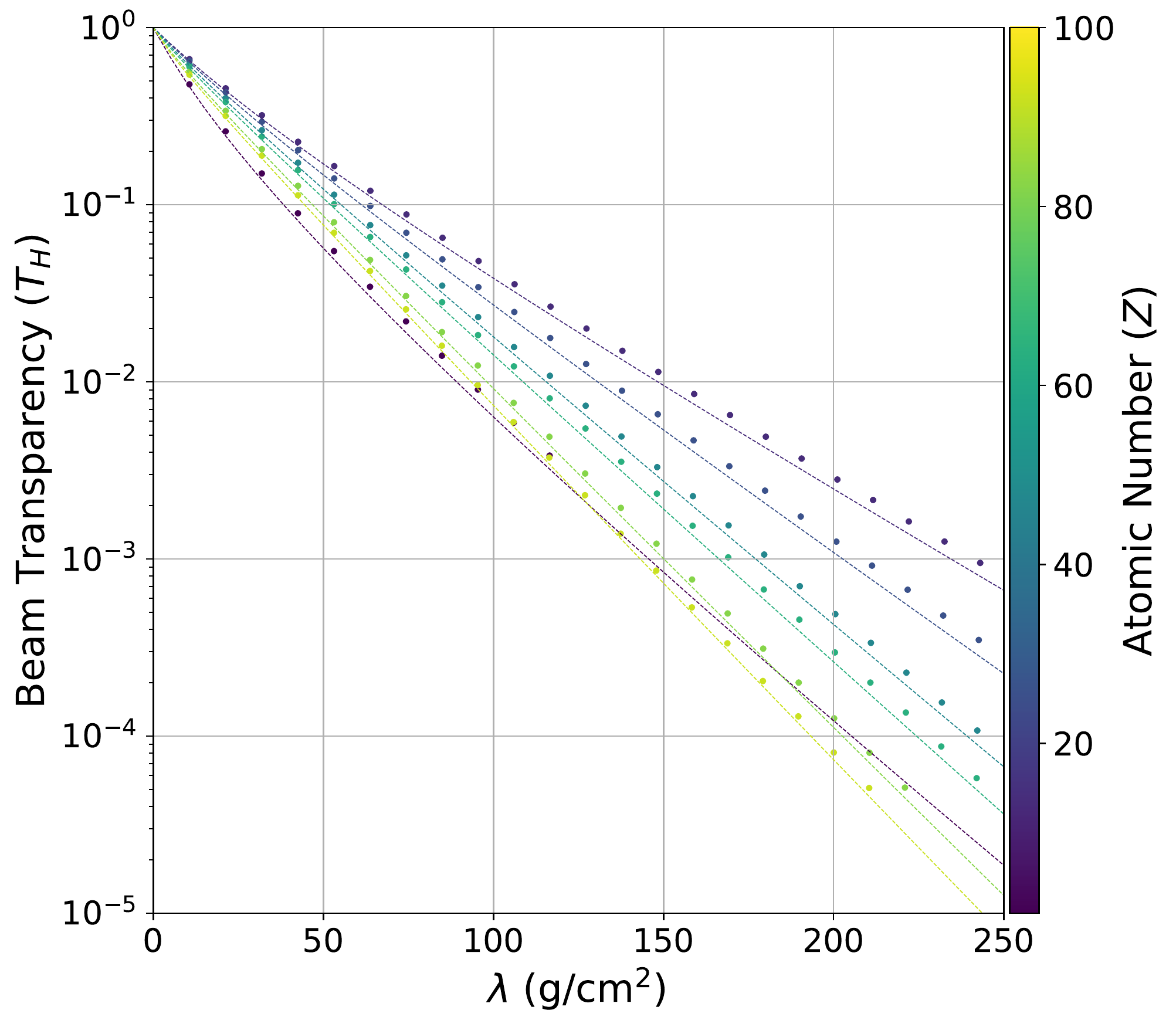}
        \caption{Comparing the free streaming model (dashed lines, Eq. \ref{freestreaming}) to transparency simulations (points), indicating noticeable discrepancies.}
        \label{T_H_lambda_freestreaming}
    \end{subfigure}
    \hfill
    \begin{subfigure}[t]{0.48\textwidth}
        \centering 
        \includegraphics[width=\textwidth]{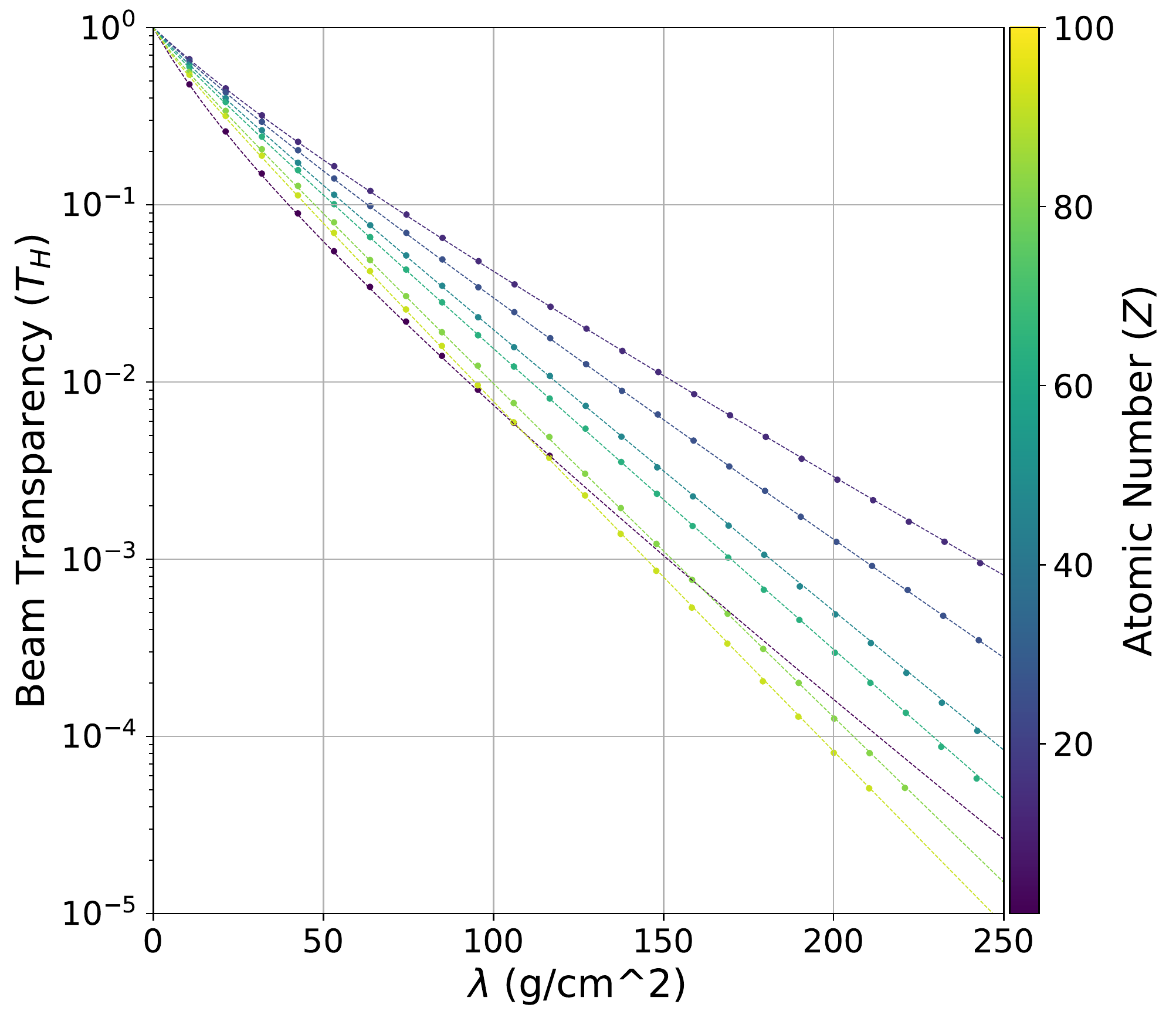}
        \caption{Comparing the semiempirical model (dashed lines, Eq. \ref{semiempirical}) to transparency simulations (points), indicating strong agreement.}
        \label{T_H_lambda_semiempirical}
    \end{subfigure}
    \caption{Photon beam transparency as a function of area density for different materials using the high energy $10$ MeV bremsstrahlung beam. The semiempirical model shows significantly reduced bias compared to the free streaming model.}
    \label{T_H_lambda}
\end{figure}

The simulated transparency measurements are compared to the semiempirical model predictions in Fig. \ref{T_H_lambda_semiempirical}, finding that the semiempirical formulation reduces the model bias by a factor of ${\approx} 10$ over the free streaming model. Fig. \ref{alpha_curve_semiempirical} plots the model predictions on an $\alpha$-curve, revealing that the semiempircial model shows improved agreement with the simulations at high-$Z$. In order to calculate the atomic number reconstruction accuracy of the free streaming and the semiempirical transparency models, the $Z$ of best fit is calculated for each of the simulated materials by finding the atomic number which minimizes the squared logarithmic error between the simulated measurements and the model predictions. The results of this analysis are shown in the first three rows of table \ref{reconstructed_Z}, along with the true material $Z$. Both models yield accurate atomic number predictions for low- to med-$Z$ materials. At high-$Z$, the free streaming model becomes unstable and is unable to identify any material which matches the simulation outputs. On the contrary, the semiempirical model maintains a relatively high degree of accuracy at high-$Z$, although the solution is not always unique for reasons described in our previous works~\cite{Lalor2023}. This high-$Z$ degeneracy is a fundamental property of dual energy X-ray scanners whereby different high-$Z$ elements can yield identical transparency measurements.
 
\subsection{Uncertainty in the Source Energy Spectra and Detector Response Function}

The source energy spectra $\phi_{\{H, L\}}$ and detector response function $D(E)$ serve as implicit parameters in both the free streaming and the semiempirical transparency models (Eqs. \ref{freestreaming} and \ref{semiempirical}). These parameters are unique to each experimental system and it has thus far been assumed that they are known exactly. However, due to the angular dependence of linac based X-ray sources, the beam spectra change at different beam angles. Furthermore, variability within the linac may result in photon beam spectra which differ from an idealized model. Similarly, the true detector response of a real system may differ from simulation output. As a result, it is not reliable to assume the source energy spectra and detector response function are all known exactly.

To probe this effect, we intentionally introduce a discrepancy between the implicit parameters used during transparency calculations and those used during transparency simulations. This procedure will enable us to benchmark the effectiveness of the semiempirical transparency model in a realistic setting in which the model parameters slightly differ from the experiment. During transparency simulations, we will continue to use $\{10, 6\}$ MeV bremsstrahlung beam spectra and record the true detector response. However, during transparency calculations, we use bremmstrahlung beam spectra with endpoint energies of $9.7 \MeV$ and $6.3 \MeV$ and the following approximate formulation of the detector response function:

\begin{equation}
D(E) \approx E \left( 1 - e^{-\mu_\detector(E) \lambda_\detector} \right) \frac{\mu_\detector^\en(E)}{\mu_\detector(E)}
\end{equation}

where $\lambda_\detector$ is the area density of the detector, $\mu_\detector$ is the mass attenuation coefficient of the detector, and $\mu_\detector^\en$ is the mass energy absorption coefficient of the detector~\cite{NIST}. We use the label ``mismatched" to distinguish instances with an intentional shift between the model parameters and the simulation parameters. It is expected that a mismatched beam spectra and detector response will worsen the atomic number reconstruction capabilities of the free streaming model. It would be useful if the semiempirical model were robust to this discrepancy by capturing these effects in the calibration procedure. In other words, perhaps through a particular selection of $a$, $b$, and $c$, it may be possible to rescale different components of the mass attenuation coefficient to counteract the mismatched model parameters. For instance, by choosing $c_L < 1$, we can artificially decrease the number of pair production interactions in our calculations, which may allow for the $6.3$ MeV semiempirical transparency model to match the results of the $6$ MeV simulations.

\begin{figure}
    \centering
    \begin{subfigure}[b]{0.48\textwidth}
        \centering
        \includegraphics[width=\textwidth]{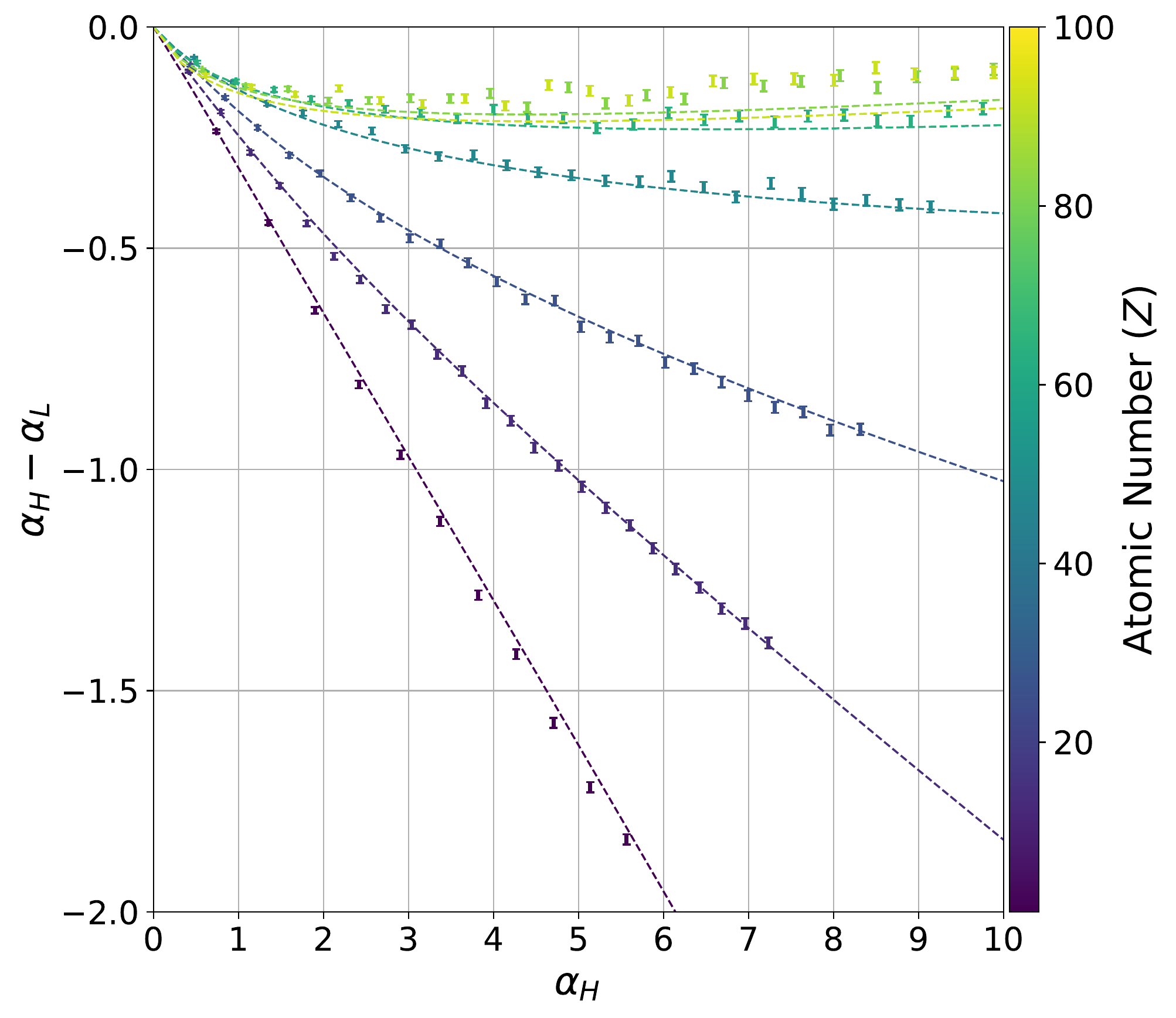}
        \caption{Comparing the free streaming model (dashed lines, Eq. \ref{freestreaming}) to transparency simulations (errorbars), indicating a discrepancy at high-$Z$.}
        \label{alpha_curve_freestreaming}
    \end{subfigure}
    \hfill
    \begin{subfigure}[b]{0.48\textwidth}  
        \centering 
        \includegraphics[width=\textwidth]{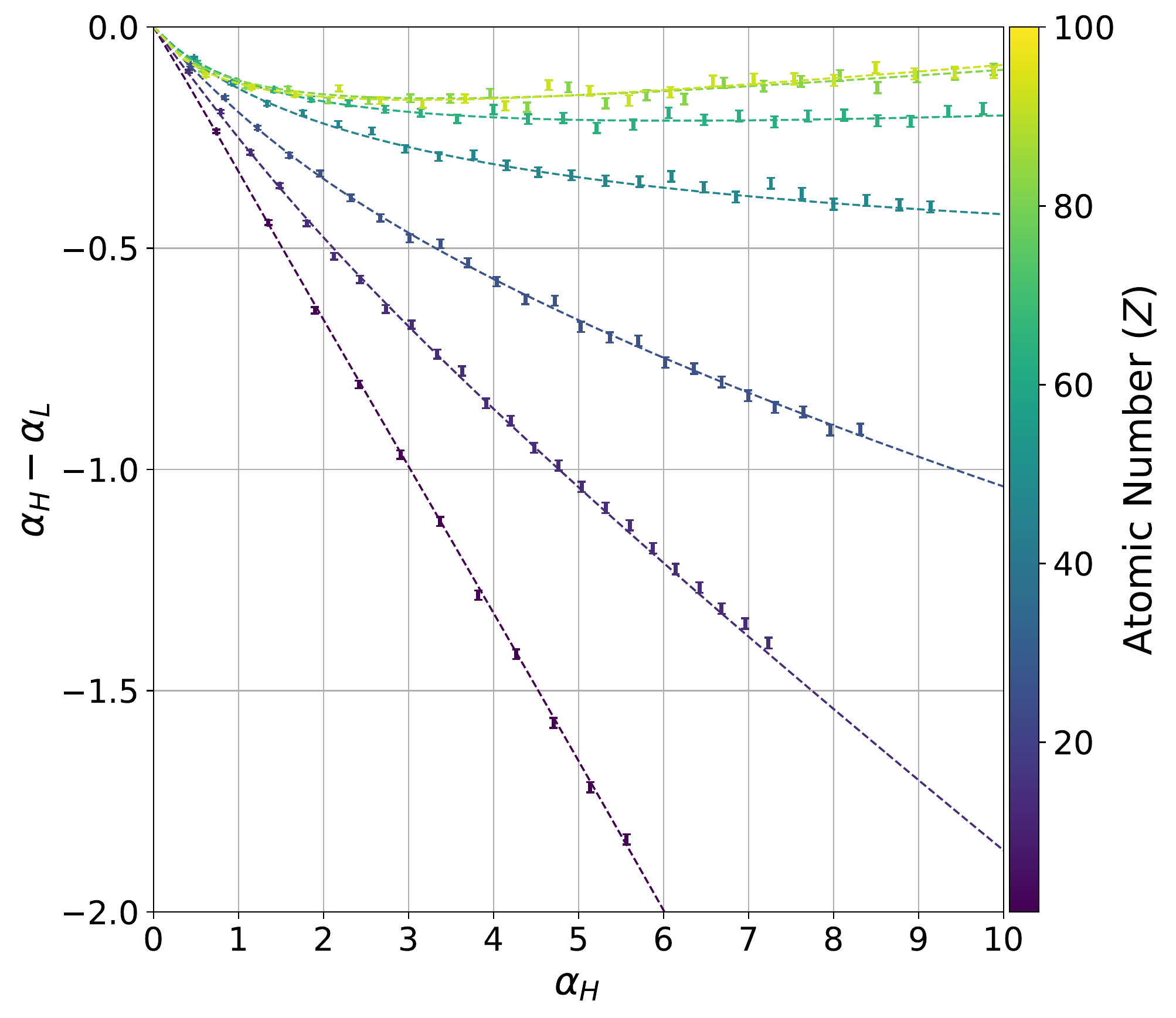}
        \caption{Comparing the semiempirical model (dashed lines, Eq. \ref{semiempirical}) to transparency simulations (errorbars), indicating strong agreement.}
        \label{alpha_curve_semiempirical}
    \end{subfigure}
    \vskip\baselineskip
    \begin{subfigure}[b]{0.48\textwidth}   
        \centering 
        \includegraphics[width=\textwidth]{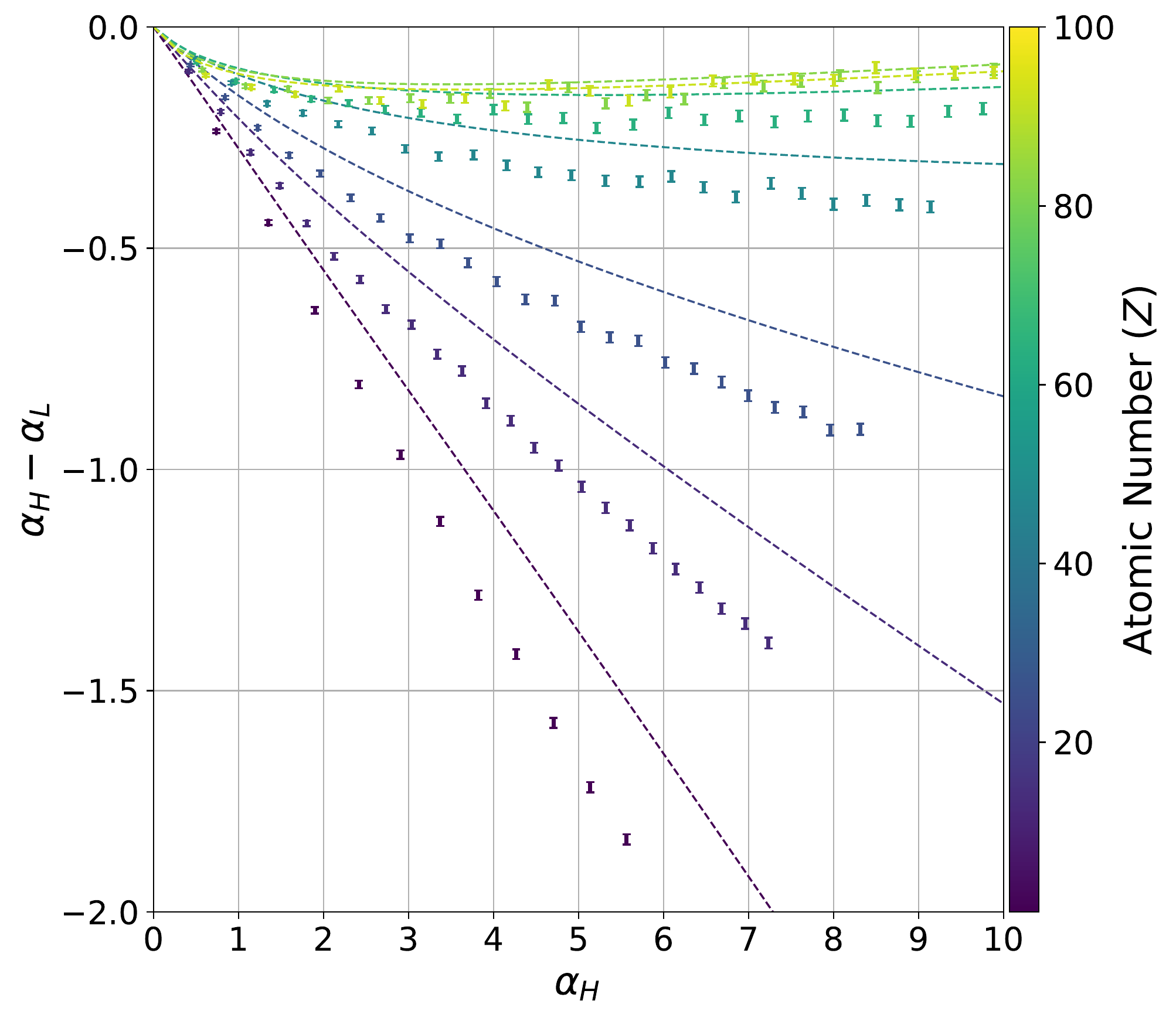}
        \caption{Comparing the free streaming model (dashed lines, Eq. \ref{freestreaming}) to transparency simulations (errorbars) using mismatched model parameters, indicating large discrepancies.}
        \label{alpha_curve_freestreaming_mismatched}
    \end{subfigure}
    \hfill
    \begin{subfigure}[b]{0.48\textwidth}   
        \centering 
        \includegraphics[width=\textwidth]{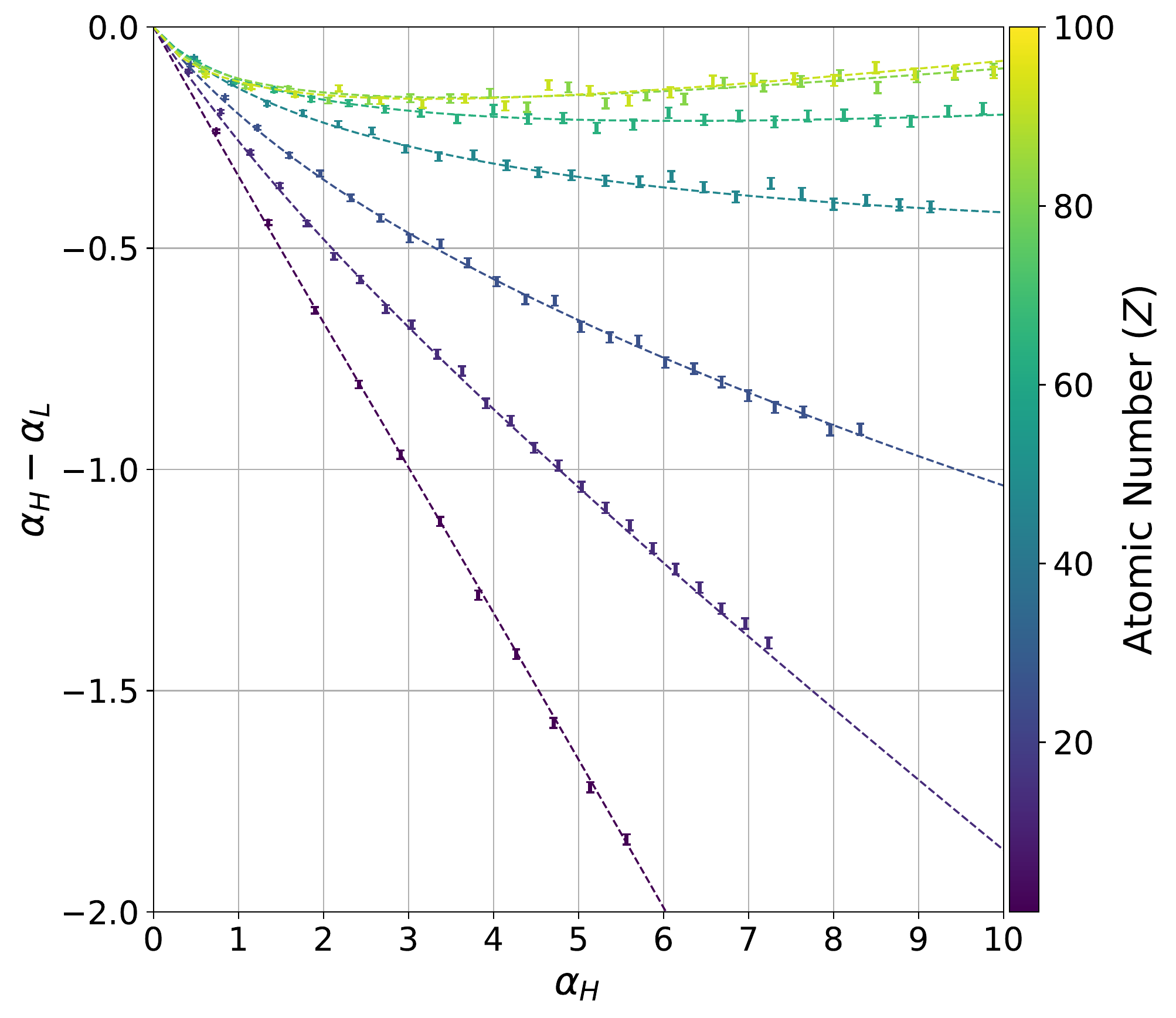}
        \caption{Comparing the semiempirical model (dashed lines, Eq. \ref{semiempirical}) to transparency simulations (errorbars) using mismatched model parameters, indicating strong agreement.}
        \label{alpha_curve_semiempirical_mismatched}
    \end{subfigure}
    \caption{$\alpha$-curves for the analysis cases presented in section \ref{Analysis}. Different materials correspond to different $\alpha$-lines, and the separation between these $\alpha$-lines enables atomic number discrimination. Compared to the free streaming model, the semiempirical model shows significantly improved agreement with the simulated data, especially when using mismatched model parameters. Materials shown are hydrogen $(Z=1)$, aluminum $(Z=13)$, iron $(Z=26)$, silver $(Z=47)$, gadolinium $(Z=64)$, lead $(Z=82)$, and uranium $(Z=92)$.}
    \label{alpha_curve}
\end{figure}

The analysis performed earlier in this section was then repeated under these new considerations. The new values of the calibration parameters $a$, $b$, and $c$ are shown in the last two rows of table \ref{abc_fit}. The subsequent transparency calculations using both the free streaming model and the semiempirical model were compared to the simulated dataset. Fig. \ref{alpha_curve_freestreaming_mismatched} reveals that using mismatched model parameters significantly worsens the accuracy of the free streaming model. On the contrary, Fig. \ref{alpha_curve_semiempirical_mismatched} demonstrates that the semiempirical transparency model continues to show excellent agreement with the simulated transparency measurements. These results are quantified in the last two rows of table \ref{reconstructed_Z}, where the semiempirical model shows significantly improved atomic number reconstruction capabilities compared to the free streaming model. As before, the solution becomes non-unique at high-$Z$ due to a fundamental and unavoidable ambiguity between heavy metals~\cite{Lalor2023}. This result shows promise for the semiempirical transparency model to enable accurate material discrimination capabilities in realistic cargo scanning applications.

\begin{table}
\begin{centering}
\begin{tabular}{c c c c}
\toprule
Bremsstrahlung endpoint energy & a & b & c \\
\midrule
10 MeV & 1.1158 & 0.9616 & 0.9973 \\
6 MeV & 0.9427 & 0.9687 & 0.9984 \\
10 MeV (simulation), 9.7 MeV (model) & 1.0659  & 0.9788 & 1.0031 \\
6 MeV (simulation), 6.3 MeV (model) & 1.1224 & 1.0343 & 0.9089 \\
\bottomrule
\end{tabular}
\caption{Calibration parameters $a$, $b$, and $c$ for each of the analysis cases considered in this study. For the first two rows, the beam spectra and detector response used in the model exactly match the simulations. For the last two rows, there is an intentional mismatch between the beam spectra and detector response used in the model compared to the simulations.}
\label{abc_fit} 
\end{centering}
\end{table}

\begin{table}
\begin{centering}
\begin{tabular}{c c c c c c c c c c c c c c}
\toprule
 & H & C & Al & Ca & Fe & Ge & Zr & Ag & Cs & Gd & W & Pb & U \\
\midrule
 True $Z$ & 1 & 6 & 13 & 20 & 26 & 32 & 40 & 47 & 55 & 64 & 74 & 82 & 92 \\
 Free streaming $Z$ & 1 & 6 & 13 & 20 & 26 & 32 & 40 & 48 & 57 & 69, 91 & N/A & N/A & N/A \\
 Semiempirical $Z$ & 1 & 6 & 13 & 20 & 26 & 32 & 41 & 48 & 56 & 64 & 76, 99 & 81, 95 & 90 \\
 \makecell{Free Steaming $Z$ \\ (mismatched)} & N/A & 1 & 8 & 15 & 20 & 26 & 34 & 40 & 47 & 55 & 64, 98 & 67, 96 & 69, 94 \\
 \makecell{Semiempirical $Z$ \\ (mismatched)} & 1 & 6 & 13 & 20 & 26 & 32 & 41 & 48 & 56 & 64 & 76, 100 & 80, 97 & 85, 94 \\
\bottomrule
\end{tabular}
\caption{Comparing the true material atomic number to the reconstructed atomic number using the free streaming transparency model (Eq. \ref{freestreaming}) and the semiempirical transparency model (Eq. \ref{semiempirical}). In the second and third rows, the models and the simulations use identical photon beam spectra and detector response functions. In the last two rows, the models use mismatched beam spectra and detector responses. Cells containing two entries indicate both materials match the simulations, whereas cells containing ``N/A" indicate no material was found to reproduce the simulated measurements.}
\label{reconstructed_Z}
\end{centering}
\end{table}





\section{Conclusion}

This work presents a semiempirical transparency model for cargo radiography measurements. The model introduces three calibration parameters which scale different components of the mass attenuation coefficient. The semiempirical model is highly flexible in its ability to be applied to different system geometries by recalculating the calibration parameters. Compared to analytic models, the semiempirical model shows significantly less bias and improved atomic number reconstruction accuracy for high-$Z$ elements. Furthermore, the semiempirical model maintains strong agreement with simulated data even when the source energy spectra and detector response function are not known exactly. Compared to empirical methods, the semiempirical model requires significantly less calibration data and shows accurate extrapolation to elements and thicknesses which were not included in the calibration step. For these reasons, the semiempirical transparency could enable improved atomic number reconstruction capabilities for dual energy cargo radiography applications. Future research should apply the semiempirical transparency model to real transparency data taken by dual energy systems to verify the accuracy in a practical setting.

\section{Acknowledgements}
This work was supported by the Department of Energy Computational Science Graduate Fellowship (DOE CSGF) under grant DE-SC0020347. The authors acknowledge advice from Brian Henderson for his expertise on the technical nuances of dual energy radiography systems. The authors declare no conflict of interest.

\newpage


\bibliography{References.bib}


\newpage

\appendix

\section{}
\label{appendix}

\subsection{Simulation Details}
\label{Simulation Details}

\begin{figure}
\begin{centering}
\includegraphics[width=0.49\textwidth]{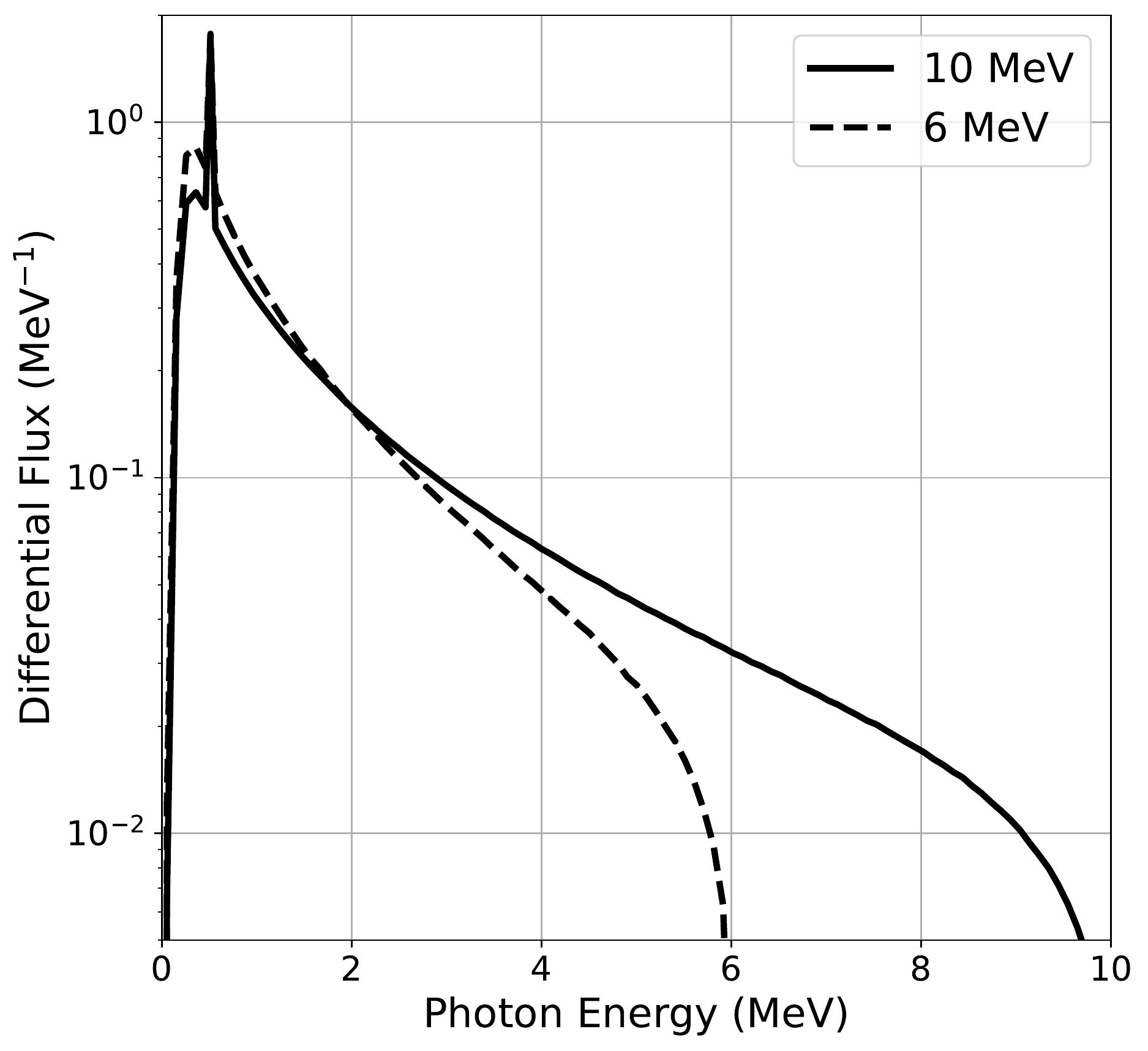}
\caption{Simulated $10$ and $6$ MeV endpoint energy bremsstrahlung beam spectra ($\phi_H$ and $\phi_L$, respectively).}
\label{spectra}
\end{centering}
\end{figure}

Simulations of the incident beam spectra $\phi_{\{H, L\}}$ and detector response matrix $R(E, E_\dep)$ were based on the work of Henderson~\cite{Henderson2019} and designed to be simple, generalizable, and representative of typical cargo scanners. In order to model the bremsstrahlung beam spectra, electrons were directed at a 0.1cm tungsten radiator backed by 1cm of copper. The bremsstrahlung photons were then recorded by a tally surface $(r = 0.2\cm)$ placed 10cm behind the tungsten target, subtending a half-angle of 20 milliradians $(\approx 1.15^{\circ})$. The resulting dual energy beam spectra are shown in Fig. \ref{spectra}. The detector response matrix was calculated by directing photons along the long axis of a  $15.0 \times 4.6 \times 30.0$ mm cadmium tungstate (CdWO$_4$) crystal. The incident energy and total deposited energy of each photon was binned to produce the detector response matrix.

\begin{figure}
\begin{centering}
\includegraphics[width=0.49\textwidth]{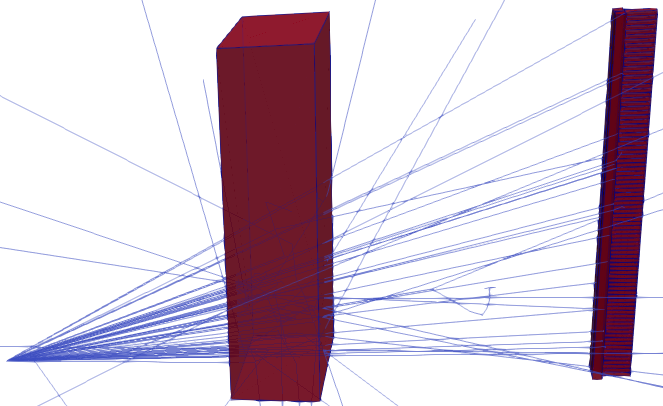}
\caption{A bremsstrahlung fan beam (left) is directed towards a target (center). A collimated stack of CdWO$_4$ detectors (right) measures the transparency of the photon beam.}
\label{simGeom}
\end{centering}
\end{figure}

To simulate transparency measurements of a target material, photons with energy sampled from $\phi_H$ or $\phi_L$ were directed in a fan beam through a large target box. The photons were then detected by a two meter stack of CdWO$_4$ detectors with a $10$cm lead collimator to filter scattered radiation. The simulation geometry is shown in figure \ref{simGeom}. This simulation was repeated for a large number of target area densities and atomic numbers to obtain a simulation dataset.

\end{document}